\documentclass[aps,prd,reprint,onecolumn,notitlepage,groupedaddress]{revtex4-1}
\usepackage{extraipa}
\usepackage{amsmath}
\usepackage{amsfonts}
\usepackage{amssymb}

\newcommand{\etal}{\textit{et al}.}

\begin{document}

\title{On the Quantum equivalence of an antisymmetric tensor field with spontaneous Lorentz violation}

\author{Sandeep Aashish}%
\email[]{sandeepa16@iiserb.ac.in}

\author{Sukanta Panda}
\email[]{sukanta@iiserb.ac.in}

\affiliation{Department of Physics, Indian Institute of Science Education and Research, Bhopal 462066, India}

%


\date{\today}

\begin{abstract}
We present an explicit proof that a minimal model of rank-2 antisymmetric field with spontaneous Lorentz violation and a classically equivalent vector field model are also quantum equivalent, by calculating quantum effective actions of both theories. We comment on the issues encountered while checking quantum equivalence in curved spacetime.
\end{abstract}



\maketitle

\section{\label{intro}Introduction}
Antisymmetric tensor fields appear in all superstring theories and are especially relevant for studies in the low-energy limit \cite{rohm1986,ghezelbash2009}. They have been studied in the past in several contexts, including strong-weak coupling duality and phase transitions \cite{quevedo1996,olive1995,polchinski1995,siegel1980,hata1981,buchbinder1988,duff1980,
bastianelli2005a,*bastianelli2005b}. 

A study relevant to the present work was carried out by Altschul \etal \cite{altschul2010}, where spontaneous Lorentz violation with various rank-2 antisymmetric field models minimally and non minimally coupled to gravity was investigated. A remarkable feature of that study is the presence of distinctive physical features with phenomenological implications for tests of Lorentz violation, even with relatively simple antisymmetric field models with a gauge invariant kinetic term. More recently, quantisation and propagator for such theories have been studied in Refs. \cite{maluf2019,aashish2019b}. Lorentz violation is also a strong candidate signal for quantum gravity, and is part of the Standard Model Extension research program \cite{bonder2015}. Such interesting phenomenological possibilities have been a strong motivation for various works on spontaneous Lorentz violation (SLV) \cite{hernaski2016,azatov2006,kostelecky1989b,*kostelecky1998,*kostelecky2004,bluhm2005,carroll1990, jackiw1999,coleman1999,bertolami1999}.

Antisymmetric tensors, and $n-$forms in general, display interesting properties with regard to their equivalence with scalar and vector fields. For instance, in four dimensions theory of a massless 2-form field (with a gauge-invariant kinetic term) is classically equivalent to a massive nonconformal scalar field, while a massless 3-form theory does not have any physical degrees of freedom (see \cite{buchbinder2008} and references therein). Likewise, a massive rank-2 antisymmetric field is clasically equivalent to a massive vector field, and a rank-3 antisymmetric field is equivalent to massive scalar field \cite{buchbinder2008}. Such properties are useful in the analysis of degrees of freedom of these theories \cite{altschul2010}. Classical equivalence implies that the actions of two theories are equivalent. However, quantum equivalence is established at the level of effective actions, and it is in general not straightforward to check especially in curved spacetime. Moreover, classical equivalence between two theories does not necessarily carry over to the quantum level, particularly in the case of spontaneously broken Lorentz symmetry \cite{seifert2010a,aashish2019b}, and thus makes for an interesting study. 

Quantum equivalence in the context of massive rank-2 and rank-3 antisymmetric fields in curved spacetime, without SLV, was first studied by Buchbinder \etal \cite{buchbinder2008} and later confirmed in Ref. \cite{shapiro2016}. The proof of quantum equivalence in Ref. \cite{buchbinder2008} was based on the zeta-function representation of functional determinants of $p$-form Laplacians appearing in the 1-loop effective action, and identities satisfied by zeta-functions for massless case \cite{rosenberg1997,elizalde1994,hawking1977}. Quantum equivalence results from these identities generalized to the massive case. In flat spacetime though, the proof is trivial as operators appearing in the effective action reduce to d'Alembertian operators due to vanishing commutators of covariant derivatives and equivalence follows by taking into account the independent components of each field.

We consider a particularly simple but interesting model of a rank-2 antisymmetric field minimally coupled to gravity, with the simplest choice of spontaneously Lorentz violating potential \cite{altschul2010}. Its classical equivalence was studied in Ref. \cite{altschul2010} in terms of an equivalent Lagrangian consisting of a vector field $A_{\mu}$ coupled to auxiliary field $B_{\mu\nu}$ in Minkowski spacetime. However, checking the quantum equivalence of such classically equivalent theories is not straightforward, in flat as well as curved spacetime. We find that the simple structure of operators breaks down due to the presence of SLV terms. As a result, the difference of their effective actions does not vanish in Minkowski spacetime, contrary to the case without SLV. However, this does not threaten quantum equivalence due to a lack of field dependence in the effective actions, which will therefore cancel after normalization.

In curved spacetime, making a conclusive statement about quantum equivalence is a nontrivial task for the following reasons. First, directly comparing effective actions using known proper time methods as in Ref. \cite{shapiro2016} is a difficult mathematical problem. Unlike the minimal operators (of the form $g_{\mu\nu}\nabla^{\mu}\nabla^{\nu} + Q$, where $Q$ is a functional without any derivative terms) found in \cite{buchbinder2008} for instance, we encounter nonminimal operators in functional determinants of the effective action, for which finding heat kernel coefficients to evaluate the determinants is a highly nontrivial task. Second, the formal arguments made in Ref. \cite{buchbinder2008} do not apply to the present case due to the non-trivial structure of operators appearing in effective actions. We point out that a resolution to this problem lies in doing a perturbative analysis of effective actions in nearly flat spacetimes, as in Ref. \cite{aashish2019b}. Nevertheless, in this work we present a derivation of one-loop effective actions of concerned theories in operator form and demonstrate the above difficulties by writing down operator structures explicitly. We use the quantization method developed in Ref. \cite{aashish2018a} to calculate the effective actions in curved spacetime using the DeWitt-Vilkovisky's covariant effective action approach (see \cite{parker2009} for review) and St{\"u}ckelberg procedure \cite{stuckelberg1957,buchbinder2007}.

The organization of this paper is as follows. Section \ref{sec2} contains a review of the antisymmetric field Lagrangian in consideration, and a derivation of classically equivalent Lagrangian. In section \ref{sec3}, we calculate the effective action for the two classically equivalent theories. Section \ref{sec4} deals with checking their quantum equivalence and problems therein.

\section{\label{sec2}Classical action}
We consider the minimal model of a rank-2 antisymmetric tensor field, $B_{\mu\nu}$, with the simplest choice of spontaneously Lorentz violating potential \cite{altschul2010},  
\begin{eqnarray}
\label{amslv0}
\mathcal{L} = -\frac{1}{12}H_{\mu\nu\lambda}H^{\mu\nu\lambda} - \frac{1}{2}\lambda\Big(B_{\mu\nu}B^{\mu\nu} - b_{\mu\nu}b^{\mu\nu}\Big)^{2}.
\end{eqnarray}
$\lambda$ here is a massless coefficient. The first term in Eq. (\ref{amslv0}) is the gauge invariant kinetic term, where,
\begin{equation}
\label{amslv1}
H_{\mu\nu\lambda} \equiv \nabla_{\mu}B_{\nu\lambda} + \nabla_{\lambda}B_{\mu\nu} + \nabla_{\nu}B_{\lambda\mu},
\end{equation}
and the second term is responsible for spontaneous Lorentz violation, giving rise to a non-zero vacuum expectation value, 
\begin{eqnarray}
\label{amslv2}
\langle B_{\mu\nu}\rangle = b_{\mu\nu}. 
\end{eqnarray} 
$b_{\mu\nu}$ is also an antisymmetric tensor, which in general may not have a simple structure, but it is possible to transform to a special observer frame in which $b_{\mu\nu}$ has a block-diagonal form with its components being real numbers, provided that $b_{\mu\nu}b^{\mu\nu}$ is nonzero \cite{altschul2010}.

It is clear from Eq. (\ref{amslv0}) that the potential contains self-interaction terms for  $B_{\mu\nu}$. Although it would be interesting to investigate quantum corrections in such a theory, it is out of scope of the current work. For the present study we are interested in the quantum properties of this theory with upto quadratic order terms in $B_{\mu\nu}$, and thus it is relevant to consider fluctuations of $B_{\mu\nu}$ around its vacuum expectation value $b_{\mu\nu}$ so that all higher order terms, including self-interaction terms can be ignored. We define the fluctuations $\crtilde{B}_{\mu\nu}$ as,
\begin{eqnarray}
\label{amslv3}
\crtilde{B}_{\mu\nu} = B_{\mu\nu} - b_{\mu\nu}.
\end{eqnarray}
Substituting Eq. (\ref{amslv3}) in Eq. (\ref{amslv0}) and neglecting higher order terms and constants, yields,
\begin{eqnarray}
\label{amslv4}
\mathcal{L} = -\frac{1}{12}\crtilde{H}_{\mu\nu\lambda}\crtilde{H}^{\mu\nu\lambda} - 2\lambda\Big(b_{\mu\nu}\crtilde{B}^{\mu\nu}\Big)^{2}.
\end{eqnarray}
where $\crtilde{H}_{\mu\nu\lambda}$ is now defined in terms of fluctuations $\crtilde{B}_{\mu\nu}$. 
For convenience, we define
\begin{eqnarray}
\label{amslv5}
b_{\mu\nu} = b n_{\mu\nu},
\end{eqnarray}
where $n_{\mu\nu}$ is an antisymmetric tensor satisfying $n_{\mu\nu}n^{\mu\nu} = 1$ so that,
\begin{eqnarray}
\label{amslv6}
b_{\mu\nu}b^{\mu\nu} = b^{2}.
\end{eqnarray}
Using Eq. (\ref{amslv5}), Lagrangian (\ref{amslv4}) can be written in a convenient form,
\begin{eqnarray}
\label{amslv7}
\mathcal{L} = -\frac{1}{12}\crtilde{H}_{\mu\nu\lambda}\crtilde{H}^{\mu\nu\lambda} - \dfrac{1}{4}\alpha^{2} \Big(n_{\mu\nu}\crtilde{B}^{\mu\nu}\Big)^{2}.
\end{eqnarray}
where $\alpha\equiv 8\lambda b^{2}$ is now a massive coefficient.

Our intention is to check the quantum equivalence of theory (\ref{amslv7}) with a classically equivalent vector theory. Classical equivalence here means equivalence at the level of Lagrangian, that is, one Lagrangian can be obtained from other and vice versa, after manipulations. Knowledge of equivalence is quite useful in analysing  the degrees of freedom and propagating modes of theories, as documented in Ref. \cite{altschul2010}. In the case $\alpha=0$ in Eq. (\ref{amslv7}), $B_{\mu\nu}$ is known to be equivalent to a scalar field $\phi$ with Lagrangian $L = -\partial_{\mu}\phi\partial^{\mu}\phi/2$ \cite{altschul2010}. For $\alpha\neq 0$, an equivalent Lagrangian can be obtained by introducing a vector field $A_{\mu}$ along with the field strength and its dual defined as,
\begin{eqnarray}
\label{aceq0}
F_{\mu\nu} = \nabla_{\mu}A_{\nu} - \nabla_{\nu}A_{\mu}, \nonumber \\
\mathcal{F}_{\mu\nu} = \dfrac{1}{2}\epsilon_{\mu\nu\rho\sigma}F^{\mu\nu}, 
\end{eqnarray}
such that, Lagrangian (\ref{amslv7}) is equivalent to \cite{altschul2010},
\begin{eqnarray}
\label{aceq1}
\mathcal{L} = \dfrac{1}{2}\crtilde{B}_{\mu\nu}\mathcal{F}^{\mu\nu} - \frac{1}{2}A^{\mu}A_{\mu} - \dfrac{1}{4}\alpha^{2} \Big(n_{\mu\nu}\crtilde{B}^{\mu\nu}\Big)^{2}.
\end{eqnarray}
For present purposes, we want to get rid of $\crtilde{B}_{\mu\nu}$ entirely in favour of a Lagrangian described solely by a vector field, and thus it is handy to make use of projections of a tensor along and transverse to $n_{\mu\nu}$,
\begin{eqnarray}
\label{aceq2}
T_{||\mu\nu} = n_{\rho\sigma}T^{\rho\sigma} n_{\mu\nu}, \nonumber \\
T_{\perp\mu\nu} = T_{\mu\nu} - T_{||\mu\nu},
\end{eqnarray}
Substituting Eq. (\ref{aceq2}) in Eq. (\ref{aceq1}), the Lagrangian density becomes,
\begin{eqnarray}
\label{aceq3}
\mathcal{L} =  \dfrac{1}{2}\crtilde{B}_{\perp\mu\nu}\mathcal{F}_{\perp}^{\mu\nu} + \dfrac{1}{2}\crtilde{B}_{||\mu\nu}\mathcal{F}_{||}^{\mu\nu} - \frac{1}{2}A^{\mu}A_{\mu} - \dfrac{1}{4}\alpha^{2} \crtilde{B}_{||\mu\nu}\crtilde{B}_{||}^{\mu\nu}.
\end{eqnarray}
Using the equations of motion of $\crtilde{B}_{||\mu\nu}$ and $\crtilde{B}_{\perp\mu\nu}$ in (\ref{aceq3}) allows us to write,
\begin{eqnarray}
\label{aceq4}
\alpha^{2}\mathcal{L} &=& \dfrac{1}{4}\mathcal{F}_{||\mu\nu}\mathcal{F}_{||}^{\mu\nu} - \frac{1}{2}\alpha^{2} A^{\mu}A_{\mu} \nonumber \\
&=& \dfrac{1}{4}\left(n_{\mu\nu}\mathcal{F}^{\mu\nu}\right)^{2}  - \frac{1}{2}\alpha^{2} A^{\mu}A_{\mu}.
\end{eqnarray}
Note that Eq. (\ref{aceq4}) incorporates the condition $\mathcal{F}_{\perp}^{\mu\nu} \approx 0$, i.e. all modes orthogonal to $n_{\mu\nu}$ are non-propagating; at the same time, the term proportional to $A_{\mu}A^{\mu}$ generates the mass term for modes along $n_{\mu\nu}$ \cite{altschul2010}. Introducing the dual of $n_{\mu\nu}$, given by $\tilde{n}_{\mu\nu} = \dfrac{1}{2}\epsilon_{\mu\nu\rho\sigma}n^{\rho\sigma}$, the classically equivalent Lagrangian in terms of $F_{\mu\nu}$ reads,
\begin{eqnarray}
\label{aceq5}
\alpha^{2}\mathcal{L} &=& \dfrac{1}{4}\left(\tilde{n}_{\mu\nu}F^{\mu\nu}\right)^{2} - \frac{1}{2}\alpha^{2} A^{\mu}A_{\mu}
\end{eqnarray}
A distinctive feature of Lagrangian (\ref{aceq5}) when compared to a generic massive vector field Lagrangian like the Proca model, is its peculiar kinetic term. Infact, kinetic terms with a Lorentz violating factor have been explored in past literatures in the context of Chern-Simons modification to Maxwell theory and alternatives to Higgs mechanism \cite{chung1999,jean2011}. Moreover, the sign of kinetic term in (\ref{aceq5}) is opposite to that in Proca model. In the context of SLV, another noteworthy feature of Lagrangian (\ref{aceq5}) is that the potential term is not affected by $n_{\mu\nu}$ unlike other vector models with SLV, for instance the Bumblebee model. It will be observed in later sections that these features lead to an effective action that has a structure different from the corresponding effective action for Lagrangian (\ref{amslv7}).

\section{\label{sec3}The Effective Action}
The classical analysis of the previous section did not take into account the gauge symmetries of equivalent Lagrangians (\ref{amslv7}) and (\ref{aceq5}). While these Lagrangians are technically not gauge invariant, they belong to a class of theories having a softly broken gauge symmetry: the kinetic terms of Lagrangians (\ref{amslv7}) and (\ref{aceq5}) are invariant under the transformations $\crtilde{B}^{\mu\nu} \longrightarrow \crtilde{B}^{\mu\nu} + \nabla_{\mu}\xi_{\nu} - \nabla_{\nu}\xi_{\mu}$ and $A_{\mu}\longrightarrow A_{\mu} + \nabla_{\mu}\Lambda$, respectively, but the potential terms are not. A standard approach for quantization of these theories is to employ the St{\"u}ckelberg procedure \cite{stuckelberg1957,buchbinder2007}. 

We first consider the Lagrangian (\ref{amslv7}). The first step is to restore the softly broken gauge symmetry through the introduction of a St{\"u}ckelberg field \cite{stuckelberg1957} $C_{\mu}$ such that the Lagrangian,
\begin{eqnarray}
\label{aeaa0}
\mathcal{L} = -\frac{1}{12}\crtilde{H}_{\mu\nu\lambda}\crtilde{H}^{\mu\nu\lambda} - \frac{1}{4}\alpha^{2}\Big[n_{\mu\nu}\Big(\crtilde{B}^{\mu\nu} + \frac{1}{\alpha}F^{\mu\nu}[C]\Big)\Big]^{2},
\end{eqnarray}
becomes gauge invariant (here, $F_{\mu\nu}\equiv\partial_{\mu}C_{\nu}-\partial_{\nu}C_{\mu}$), and reduces to original Lagrangian (\ref{amslv7}) in the gauge $C_{\mu}=0$. The new Lagrangian (\ref{aeaa0}) is invariant under the symmetries,
\begin{eqnarray}
\label{aeaa1}
\crtilde{B}^{\mu\nu} &\longrightarrow & \crtilde{B}^{\mu\nu} + \nabla_{\mu}\xi_{\nu} - \nabla_{\nu}\xi_{\mu}, \nonumber \\
C_{\mu} &\longrightarrow & C_{\mu} - \alpha\xi_{\mu},
\end{eqnarray}
and, 
\begin{eqnarray}
\label{aeaa2}
C_{\mu} &\longrightarrow & C_{\mu} + \nabla_{\mu}\Lambda , \nonumber \\
\crtilde{B}^{\mu\nu} &\longrightarrow & \crtilde{B}^{\mu\nu} .
\end{eqnarray}
In addition to the above symmetries of fields, there exists a set of transformation of gauge parameters $\Lambda$ and $\xi_{\mu}$ that leaves the fields $B_{\mu\nu}$ and $C_{\mu}$ invariant,
\begin{eqnarray}
\label{aeaa3}
\xi_{\mu} &\longrightarrow & \xi_{\mu} + \nabla_{\mu}\psi , \nonumber \\
\Lambda &\longrightarrow & \Lambda + \alpha\psi .
\end{eqnarray}

Now, the gauge fixing procedure requires that a gauge condition be chosen for each of the fields $B_{\mu\nu}$ and $C_{\mu}$ as well as for the parameter $\xi_{\mu}$, so that the redundant degrees of freedom due to symmetries (\ref{aeaa1}), (\ref{aeaa2}) and (\ref{aeaa3}) are taken care of. An important consideration while choosing a gauge condition is to ensure that all cross terms of fields in the Lagrangian cancel out or lead to a total derivative term, so that path integral can be computed with ease. Keeping this in mind, we choose the gauge condition for $B_{\mu\nu}$ to be (a similar choice for gauge condition in the context of Bumblebee model was considered in Ref. \cite{escobar2017})
\begin{eqnarray}
\label{aeaa6}
\chi_{\xi_{\nu}} = n_{\mu\nu}n_{\rho\sigma}\nabla^{\mu}\crtilde{B}^{\rho\sigma} + \alpha C_{\nu}.
\end{eqnarray}
It turns out that the gauge fixing action term corresponding to Eq. (\ref{aeaa6}) introduces yet another soft symmetry breaking in $C_{\mu}$ \citep{aashish2018a}, so one has to introduce another St{\"u}ckelberg field $\Phi$ so that, 
\begin{eqnarray}
\label{aeaa7}
C_{\mu} \longrightarrow C_{\mu} + \dfrac{1}{\alpha}\nabla_{\mu}\Phi.
\end{eqnarray}
This modifies the symmetry in Eq. (\ref{aeaa2}) by an additional shift transformation, 
\begin{eqnarray}
\label{aeaa8}
\Phi \longrightarrow \Phi - \alpha\Lambda.
\end{eqnarray}
From Eqs. (\ref{aeaa2}) and (\ref{aeaa8}), the gauge condition for $C_{\mu}$ can be chosen to be,
\begin{eqnarray}
\label{aeaa9}
\chi_{\Lambda} = \nabla^{\mu}C_{\mu} + \alpha\Phi.
\end{eqnarray}
Similarly, for the symmetry of parameters, Eq. (\ref{aeaa3}), we choose 
\begin{eqnarray}
\label{beaa0}
\check{\chi}_{\psi} = \nabla^{\mu}\xi_{\mu} - \alpha\Lambda.
\end{eqnarray}
The gauge conditions chosen above are incorporated in the action for (\ref{aeaa0}) through ``gauge-fixing Lagrangian" terms of the form $-\frac{1}{2}\chi_{(\cdot)}^{2}$ for each of the conditions (\ref{aeaa6}), (\ref{aeaa9}) and (\ref{beaa0}). The total gauge fixed Lagrangian is given by 
\begin{eqnarray}
\label{beaa1}
\mathcal{L}_{2}^{GF} =  -\frac{1}{12}\crtilde{H}_{\mu\nu\lambda}\crtilde{H}^{\mu\nu\lambda} - \frac{1}{4}\alpha^{2}\Big(n_{\mu\nu}\crtilde{B}^{\mu\nu}\Big)^{2} - \dfrac{1}{4}\Big(n_{\mu\nu}F^{\mu\nu}\Big)^{2} - \frac{1}{2}\Big(n_{\mu\nu}n_{\rho\sigma}\nabla^{\mu}\crtilde{B}^{\rho\sigma}\Big)^{2} \nonumber \\ - \frac{1}{2}\alpha^{2}C_{\nu}C^{\nu} - \frac{1}{2}(\nabla_{\mu}\Phi)^{2} - \frac{1}{2}(\nabla^{\mu}C_{\mu})^{2} - \frac{1}{2}\alpha^{2}\Phi^{2}.
\end{eqnarray}
Following the method developed in Ref. \cite{aashish2018a}, the calculation of ghost determinants proceeds as follows. We rewrite $\chi_{\xi_{\nu}}$ as,
\begin{eqnarray}
\label{beaa4}
\chi_{\xi_{\nu}}[\crtilde{B}^{\mu\nu}_{\xi_{\nu}}, C^{\mu}_{\xi_{\nu}}] = \chi_{\xi_{\nu}}[\crtilde{B}^{\mu\nu},C^{\mu},{\xi_{\nu}},\Lambda,\check{\chi}_{\psi}],
\end{eqnarray}
which yields,
\begin{eqnarray}
\label{beaa5}
\chi_{\xi_{\nu}} = n_{\mu\nu}n_{\rho\sigma}\nabla^{\mu}\crtilde{B}^{\rho\sigma} + \alpha C_{\nu}
+ 2n_{\mu\nu}n_{\rho\sigma}\nabla^{\mu}\nabla^{\rho}\xi^{\sigma} + \nabla_{\nu}\nabla_{\mu}\xi^{\mu} - \alpha^{2}\xi_{\nu} - \nabla_{\nu}\check{\chi}_{\psi}.
\end{eqnarray}
Then, using the definition of $Q'^{\xi_{\mu}}_{\xi_{\nu}}$, we get
\begin{eqnarray}
\label{beaa6}
Q'^{\xi_{\nu}}_{\xi_{\alpha}} = \left(\dfrac{\delta\chi_{\xi_{\nu}}}{\delta\xi_{\alpha}}\right)_{\xi_{\mu} = 0} &=& 2n_{\mu\nu}n_{\rho\alpha}\nabla^{\mu}\nabla^{\rho} + \nabla_{\nu}\nabla_{\alpha} - \alpha^{2}\delta_{\nu\alpha}.
\end{eqnarray}
A straightforward calculation leads to other non-zero components of ghost determinant,
\begin{eqnarray}
\label{beaa7}
Q'^{\Lambda}_{\Lambda} = \dfrac{\delta\chi_{\Lambda}}{\delta\Lambda} = \Box_{x} - \alpha^{2} \\
\label{beaa8}
\check{Q}^{\psi}_{\psi} \equiv \dfrac{\delta\check{\chi}_{\psi}}{\delta\psi} = \Box_{x} - \alpha^{2}
\end{eqnarray}
Using the definition of effective action obtained in \cite{aashish2018a}, 
\begin{eqnarray}
\label{beaa9}
\exp(i\Gamma[\bar{B},\bar{C}]) = \int\prod_{\mu}dC_{\mu}\prod_{\rho\sigma}dB_{\rho\sigma}\prod_{x}d\Phi
\det(Q'^{\Lambda}_{\Lambda})\det(Q'^{\xi_{\nu}}_{\xi_{\alpha}})(\det\check{Q}^{\psi}_{\psi})^{-1}\times \nonumber \\  
 \exp\left\{i \Bigg(\int d v_{x} \mathcal{L}_{2}^{GF}\Bigg)  + (\bar{B}_{\mu\nu}-B_{\mu\nu})\dfrac{\delta}{\delta \bar{B}_{\mu\nu}}\Gamma[\bar{B},\bar{C}] \right. \nonumber \\ \left. + (\bar{C}_{\mu}-C_{\mu})\dfrac{\delta}{\delta \bar{C}_{\mu}}\Gamma[\bar{B},\bar{C}] \right\},
\end{eqnarray}
The 1-loop effective action is obtained as,
\begin{eqnarray}
\label{ceaa0}
\Gamma_{2}^{(1)} = \frac{i\hbar}{2}\Big[\ln\det(\crtilde{D}_{2} - \alpha^{2}n^{\mu\nu}n_{\rho\sigma}) - \ln\det(\crtilde{D}_{1}-\alpha^{2}) + \ln\det(\Box_{x} - \alpha^{2})\Big]
\end{eqnarray}
where,
\begin{eqnarray}
\label{cea1}
\crtilde{D}_{2}{}^{\mu\nu}_{\ \ \rho\sigma}B^{\rho\sigma} &\equiv & \nabla_{\alpha}\nabla^{\alpha}B^{\mu\nu} + \nabla_{\alpha}\nabla^{\mu}B^{\nu\alpha} + \nabla_{\alpha}\nabla^{\nu}B^{\alpha\mu} + 2 n^{\mu\nu}n_{\rho\sigma}n^{\alpha\sigma}n_{\beta\gamma}\nabla^{\rho}\nabla_{\alpha}B^{\beta\gamma}, \nonumber \\
\crtilde{D}_{1}{}^{\mu}_{\ \nu}C^{\nu} &\equiv & 2 n^{\nu\mu}n_{\rho\sigma}\nabla_{\nu}\nabla^{\rho}C^{\sigma} + \nabla^{\mu}\nabla_{\nu}C^{\nu}.
\end{eqnarray}
It is to be noted that the coefficient of $\alpha^{2}$ in the first term in Eq. (\ref{ceaa0}) ensures that massive modes correspond to field components along vacuum expectation tensor $n_{\mu\nu}$ and massless modes correspond to transverse components. An interesting observation here is the last term, which is unaffected by $n_{\mu\nu}$. In case of no SLV, the last term causes the quantum discontinuity when going from massive to massless case \cite{shapiro2016}. 

To compare Eq. (\ref{ceaa0}) with the effective action of classically equivalent Lagrangian, the Lagrangian in (\ref{aceq5}) is treated with the St{\"u}ckelberg procedure to obtain,
\begin{eqnarray}
\label{ceaa1}
\tilde{\mathcal{L}}_{1} &=& \dfrac{1}{4}\Big(\tilde{n}_{\mu\nu}F^{\mu\nu}\Big)^{2} - \dfrac{1}{2}\alpha^{2}(C_{\mu} + \dfrac{1}{\alpha}\nabla_{\mu}\Phi)^{2}
\end{eqnarray}
The above Lagrangian is invariant under a transformation identical to Eqs. (\ref{aeaa2}) and (\ref{aeaa8}),
\begin{eqnarray}
\label{ceaa2}
C_{\mu}\longrightarrow C_{\mu} + \nabla_{\mu}\Lambda, \quad \Phi\longrightarrow \Phi - \alpha\Lambda.
\end{eqnarray}
With the gauge condition Eq. (\ref{aeaa9}), the gauge fixed Lagrangian reads,
\begin{eqnarray}
\label{ceaa3}
\tilde{\mathcal{L}}_{1}^{GF} = \dfrac{1}{2}C_{\mu}D_{1}C^{\mu} - \dfrac{1}{2}\alpha^{2}C_{\mu}C^{\mu} + \dfrac{1}{2}\Phi (\Box_{x} - \alpha^{2})\Phi.
\end{eqnarray}
where, 
\begin{eqnarray}
\label{deaa0}
D_{1}C_{\mu} = -2\tilde{n}_{\nu\mu}\tilde{n}_{\rho\sigma}\nabla^{\nu}\nabla^{\rho}C^{\sigma} + \nabla_{\mu}\nabla_{\nu}C^{\nu}.
\end{eqnarray}
It is straightforward to check that the 1-loop effective action is,
\begin{eqnarray}
\label{ceaa4}
\Gamma_{1}^{(1)} = \frac{i\hbar}{2}\Big[\ln\det(D_{1}-\alpha^{2}) - \ln\det(\Box_{x} - \alpha^{2})\Big].
\end{eqnarray}
Similar to Eq. (\ref{ceaa0}), the scalar term is unaffected by $n_{\mu\nu}$ and the operator $D_{1}$ possesses a non-trivial structure. The expression for $D_{1}$ has a striking resemblance to that of $\crtilde{D}_{1}$, which has opposite sign in the first term and $n_{\mu\nu}$ instead of $\tilde{n}_{\mu\nu}$. Particularly interesting is the fact that this difference is, by design, built into the equivalent Lagrangian (\ref{aceq5}) and is apparent even before, in Eq. (\ref{beaa1}), where the kinetic part of St{\"u}ckelberg field $-\frac{1}{4}(n_{\mu\nu}F^{\mu\nu})^{2}$ has a sign opposite to that of Eq. (\ref{aceq5}).

\section{\label{sec4}Quantum equivalence in flat spacetime}
To compare Eqs. (\ref{ceaa0}) and (\ref{ceaa4}), we define the difference in 1-loop effective actions  given by,
\begin{eqnarray}
\label{aqeq0}
\Delta\Gamma &=& \Gamma_{2}^{(1)} - \Gamma_{1}^{(1)} \nonumber \\
&=& \frac{i\hbar}{2}\Big[\ln\det(\crtilde{D}_{2} - \alpha^{2}n^{\mu\nu}n_{\rho\sigma}) - \ln\det(\crtilde{D}_{1}-\alpha^{2}) - \ln\det(D_{1}-\alpha^{2})\nonumber \\ 
&& + 2\ln\det(\Box_{x} - \alpha^{2})\Big].
\end{eqnarray}
In contrast, the corresponding difference in 1-loop effective action in the case of massive antisymmetric and vector fields, with mass $m$, with no spontaneous Lorentz violation is given by \cite{buchbinder2008},
\begin{eqnarray}
\label{aqeq1}
\Delta\Gamma' = \frac{i\hbar}{2}\Big[\ln\det(\Box_{2} - m^{2}) - 2\ln\det(\Box_{1}-m^{2}) + 2\ln\det(\Box_{x} - m^{2})\Big],
\end{eqnarray}
where,
\begin{eqnarray}
\label{aqeq2}
\Box_{2}B_{\mu\nu} &=& \Box_{x}B_{\mu\nu} - [\nabla^{\rho},\nabla_{\nu}]B_{\mu\rho} - [\nabla^{\rho},\nabla_{\mu}]B_{\rho\nu}, \nonumber \\
\Box_{1}C^{\mu} &=& \Box_{x}C^{\nu} - [\nabla^{\nu},\nabla_{\mu}]C^{\mu}.
\end{eqnarray}
This comparison between cases with and without SLV is quite insightful, because it helps in understanding how the functional operators change due to the presence of Lorentz violating terms. In the later case, the operator for St{\"u}ckelberg vector field and that for vector field of equivalent Lagrangian are equal, while in the former case they are not, as was noted earlier. Moreover, operators in Eq. (\ref{aqeq0}) do not contain the commutator terms due to presence of $n_{\mu\nu}$, and hence do not simplify in flat spacetime unlike their counterparts in Eq. (\ref{aqeq1}).

In flat spacetime, it can be explicitly checked that Eq. (\ref{aqeq1}) vanishes, taking into account the number of independent components of respective fields (eight, four and one for antisymmetric, vector and scalar fields respectively), because the commutators in Eq. (\ref{aqeq2}) vanish and hence the operators $\Box_{2}$, $\Box_{1}$, and $\Box_{x}$ are identical. Inferring quantum equivalence is thus trivial. However, this is clearly not the case in Eq. (\ref{aqeq0}) due to the non-trivial structure of operators $\crtilde{D}_{2}$ and $\crtilde{D}_{1}$. This can be demonstrated in a rather simple example when a special choice of tensor $n_{\mu\nu}$ is considered. It can be shown that in Minkowski spacetime, $n_{\mu\nu}$ can be chosen to have a special form
\begin{eqnarray}
\label{afse0}
n_{\mu\nu} = 
\left(\begin{matrix}
0 & -a & 0 & 0\\
a & 0 & 0 & 0\\
0 & 0 & 0 & b\\
0 & 0 & -b & 0
\end{matrix}\right),
\end{eqnarray}
where $a$ and $b$ are real numbers, provided atleast one of the quantities $x_{1}\equiv -2(a^{2}-b^{2})$ and $x_{2}\equiv 4ab$ are non-zero \citep{altschul2010}. For simplicity, and dictated by the requirements for non-trivial monopole solutions \cite{seifert2010b}, we may choose $b=0$. Further, the constraint $n_{\mu\nu}n^{\mu\nu}=1$ implies that $a=1/\sqrt{2}$. Therefore, the only non-zero components of $n_{\mu\nu}$ are $n_{10}=1/\sqrt{2}$ and $n_{01}=-1/\sqrt{2}$. For the dual tensor $\tilde{n}_{\mu\nu}$, the non-zero components are $\tilde{n}_{32}=-1/\sqrt{2}$ and $\tilde{n}_{23}=1/\sqrt{2}$. Substituting in Eqs. (\ref{cea1}) and (\ref{deaa0}), one obtains, for the non-zero components of $n_{\mu\nu}$ and $\tilde{n}_{\mu\nu}$,
\begin{eqnarray}
\label{afse1}
D_{1}C^{2} &=& \partial_{2}^{2}C^{2} - \partial_{3}^{2}C^{2} + 2\partial_{3}\partial^{2}C^{3} + \partial^{2}\partial_{i}C^{i}, \nonumber \\
D_{1}C^{3} &=& - \partial_{2}^{2}C^{3} + \partial_{3}^{2}C^{3} + 2\partial^{3}\partial_{2}C^{2} + \partial^{3}\partial_{i}C^{i}, \nonumber \\
\crtilde{D}_{1}C^{0} &=&  \partial_{0}^{2}C^{0} + \partial_{1}^{2}C^{0} + \partial_{0}\partial_{j}C^{j}, \\
\crtilde{D}_{1}C^{1} &=& \partial_{0}^{2}C^{1} + \partial_{1}^{2}C^{1} + \partial_{1}\partial_{j}C^{j} , \nonumber \\
\crtilde{D}_{2}\crtilde{B}^{10} &=& \Box_{x}\crtilde{B}^{10} + \partial_{j}\left(\partial_{1}\crtilde{B}^{0j} + \partial_{0}\crtilde{B}^{j1}\right) = - \crtilde{D}_{2}\crtilde{B}^{01}, \nonumber
\end{eqnarray}
where, $j=2,3$ and $i=0,1$. The remaining components of operators $\crtilde{D}_{2}$, $\crtilde{D}_{1}$ and $D_{1}$ are given by,
\begin{eqnarray}
\label{afse2}
\crtilde{D}_{2}\crtilde{B}^{jk} &=& \Box_{x}\crtilde{B}^{jk} + \partial_{\mu}\partial^{j}\crtilde{B}^{k\mu} + \partial_{\mu}\partial^{k}\crtilde{B}^{\mu j}, \quad \crtilde{B}^{jk}\neq \crtilde{B}^{10} \nonumber \\
\crtilde{D}_{1}C^{l} &=& \partial^{l}\partial_{\nu}C^{\nu}, \quad l=2,3 \\
D_{1}C^{k} &=& \partial^{k}\partial_{\nu}C^{\nu}, \quad k=0,1. \nonumber
\end{eqnarray}
An interesting feature here, compared to the case of Eq. (\ref{aqeq1}), is that Eqs. (\ref{afse1}) and (\ref{afse2}) substituted in Eq. (\ref{aqeq0}) show explicitly that $\Delta\Gamma$ does not vanish. However,  functional determinants in Eq. (\ref{aqeq0}) do not have field dependence and can only contribute as infinite (regularization-dependent) constants \cite{simon1977,dunne2008}. Hence, each determinant in Eq. (\ref{aqeq0}) can be normalized to identity and will thus be equal to each other. And once again, taking into account the degrees of freedom of corresponding tensor, vector and scalar fields, similar to Eq. (\ref{aqeq1}), they will cancel for all physical processes. This proves the quantum equivalence of theories (\ref{ceaa0}) and (\ref{ceaa4}) in flat spacetime. 

A check of quantum equivalence in curved spacetime is out of the scope of present work, because of the lack of mathematical tools to compute functional determinants in Eq. (\ref{aqeq0}). More specifically, to evaluate the expansions these determinants one needs the correct heat kernels for operators $\crtilde{D}_{2}$, $\crtilde{D}_{1}$ and $D_{1}$. However, a perturbative approach can be undertaken to address this issue in a nearly flat spacetime as implemented in Ref. \cite{aashish2019b}.

\section{Summary}
We derived the Lagrangian for a vector field $C_{\mu}$ which is classically equivalent to a rank-2 antisymmetric tensor field with a spontaneously Lorentz violating potential, by extending the calculations carried out in Ref. \cite{altschul2010}. The 1-loop effective action in terms of functional operators was obtained for both theories, and it was found that the operators have complicated structures due to the presence of vacuum expectation tensor $n_{\mu\nu}$. In flat spacetime, we explicitly checked for a simple choice of $n_{\mu\nu}$ that although the difference of effective actions, $\Delta\Gamma$, does not vanish, their quantum equivalence still holds for physical processes once normalization of functional determinants are taken into account. This confirms, in flat spacetime, the fact that two free field theories which are classically equivalent, must also be quantum equivalent.

In curved spacetime, however, it is difficult to make a precise statement because an explicit comparison of operators is not possible unless one uses a regularization scheme to find an appropriate expression for operators in $\Delta\Gamma$, as done in Refs. \cite{shapiro2016} and \cite{buchbinder2008}. The question of quantum equivalence is important, not only due to its usefulness in analysing the degrees of freedom, as documented in Ref. \cite{altschul2010}, but also due to the consequences to formal properties of theories. For example, Seifert \cite{seifert2010a} showed that interaction of vector and tensor theories with gravity are different when topologically non-trivial monopole-like solutions of the spontaneous symmetry breaking equations exist. Although perturbative methods can be employed in a nearly flat spacetime \cite{aashish2019b}, a good starting point for addressing this issue in a general spacetime would be to explicitly write the heat kernel for these operators. 

\begin{acknowledgments}
 This work was partially funded by DST (Govt. of India), Grant No. SERB/PHY/2017041. The authors are grateful to Prof. Alan Kostelecky for useful comments on an earlier version of this paper. 
\end{acknowledgments}

%

\bibliography{ref}

\end{document}